\documentclass[12pt]{article}
\usepackage{graphicx}
\usepackage{amsmath,amssymb,amsfonts}
\usepackage{mathrsfs}
\usepackage{latexsym}
\usepackage{bm}
\usepackage{bbm}
\usepackage{cite}
\usepackage{subfigure}
\usepackage{mathrsfs}
\usepackage{indentfirst}
\usepackage{amsmath}
\usepackage{amssymb}
\usepackage{cases}

\usepackage[colorlinks]{hyperref}
\hypersetup{citecolor=blue,linkcolor=blue,urlcolor=blue}

\begin{document}

\title{\Large\bf
Direct detection of dark matter with resonant annihilation}

\vspace{0.3truecm}
\author{
 Bo Li\footnote{Email: libo@itp.ac.cn}
  \ and  Yu-Feng Zhou\footnote{Email: yfzhou@itp.ac.cn}
  \\ \\
 \textit{State Key Laboratory of Theoretical Physics},\\
  \textit{Kavli Institute for Theoretical Physics China,}\\
  \textit{Institute of  Theoretical Physics, Chinese Academy of Sciences,}\\
  \textit{Beijing, 100190, P.~R.~China}
}
\date{\today}

\maketitle
\begin{abstract}
In the scenario where
the dark matter (DM) particles $\chi\bar\chi$ pair annihilate through a resonance particle $R$,
the constraint from  DM relic density makes
the corresponding cross section for DM-nuclei elastic scattering extremely small,
and
can be below the  neutrino background induced by
the coherent neutrino-nuclei scattering,
which makes the DM particle beyond the reach of
the conventional  DM direct detection experiments.
We present an improved  analytical calculation of
the DM relic density in the case of  resonant DM annihilation  for $s$- and $p$-wave cases and
invesitgate  the condition for the DM-nuclei scattering cross section to be
above the neutrino background.
We show that
in Higgs-portal type models,
for DM particles with $s$-wave annihilation,
the spin-independent DM-nucleus scattering cross section is proportional to
$\Gamma_{R}/m_{R}$,
the ratio of the decay width and the mass of  $R$.
For a typical DM particle mass $\sim50$ GeV, the condition leads to
$\Gamma_{R}/m_{R} \gtrsim \mathcal{O}(10^{-4})$.
In $p$-wave annihilation case,
the  spin-independent scattering cross section is insensitive to $\Gamma_{R}/m_{R}$,
and
is always above the neutrino background,
as long as the DM particle  is lighter than the top quark.
The real singlet DM model is discussed as a concrete example.
%
%
%
\end{abstract}

\newpage

\section{Introduction}
Dark matter (DM) contributes to 26.8\% of the total energy density of the Universe~\cite{Ade:2013sjv},
yet its particle nature remains largely unknown.
%
%
%
The leading candidates for  DM are
weakly interacting massive particles (WIMPs). 
WIMPs can naturally obtain the observed relic density,
and
the predicted cross sections of the WIMP-nuclei  scattering are
usually within the reach of the current DM direct detection experiments. 
In the case where
the DM annihilation cross section times the relative velocity $\sigma v_{\rm rel}$ is  a constant,
such as that in the simple $s$-wave annihilation cases,
the DM relic density can be calculated analytically using the standard freeze-out approximation.
The connection  between the DM relic density and the DM-nuclei scattering cross section can be
straightforwardly established.

However,
in many DM models and DM interaction mechanisms,
the velocity dependence of $\sigma v_{\rm rel}$ can be complicated.
For instance,
to explain both the relic density and the cosmic-ray positron excess
observed by
PAMELE~\cite{Adriani:2008zr},
Fermi-LAT~\cite{FermiLAT:2011ab},
and AMS-02~\cite{Aguilar:2013qda,Aguilar:2014mma},
the mechanism of Sommerfeld enhancement has been invoked,
which introduces a velocity-dependent DM annihilation cross section
\cite{Hisano:2002fk,Hisano:2003ec,Hisano:2004ds,
ArkaniHamed:2008qn,
Liu:2013vha,Chen:2013bi,
Liu:2011aa,Liu:2011mn
}
to account for the larger cross section required by the data
\cite{DeSimone:2013fia,Jin:2013nta,Jin:2014ica}.

In a wide class of  DM models,
the DM particle  $\chi$ can annihilate into
the standard model (SM) particles through an $s$-channel resonance particle $R$.
Such as the singlet  scalar DM models
\cite{Silveira:1985rk,McDonald:1993ex,Burgess:2000yq,Davoudiasl:2004be,He:2009yd,Gonderinger:2009jp,Mambrini:2011ik,Cline:2012hg,Cline:2013gha,Guo:2010hq}
,
the left-right symmetric models with extended
stable scalar sectors
\cite{Wu:2007kt,Wu:2007gb,%
Guo:2008si,Guo:2010vy,Guo:2010sy,Guo:2011zze,%
Liu:2012bm,Bao:2011nh
}
and the fermionic DM models
\cite{
Kim:2006af,Baek:2011aa,LopezHonorez:2012kv,Esch:2013rta,%
Li:2014wia,Bao:2013zua,Zhou:2011fr
}.
The presence of the scalar $R$ can also play an important role in
electroweak phase transion
\cite{Cline:2012hg,Espinosa:2011ax,Chung:2012vg,Choi:1993cv,Ham:2004cf,Ahriche:2007jp,Profumo:2007wc,Cline:2009sn,Espinosa:2011eu,Fairbairn:2013uta,Li:2014wia}
and modify the interpretation of the DM-nuclei scattering
\cite{Jin:2012jn,Li:2014vza}.

Near the resonance point the kinetic energy of the DM particles is non-negligible,
which makes $\sigma v_{\rm rel}$ velocity dependent,
and
leads to the enhancement of DM annihilation cross section at lower tempertures,
the so called Breit-Wigner enhancement\cite{Ibe:2008ye,Guo:2009aj}.
In the scenario of resonant dark matter (DM) annihilation,
under the constraint of  DM relic density,
the cross section for DM-nuclei elastic scattering can be extremely small
such that
it can fall below the  background induced by the coherent neutrino-nuclei scattering,
which make it undetectable by the current DM direct detection technology.
It is of importance to know under what condition this phenomena will occur.
In order to establish the correlation between the DM relic density and
the DM-nuclei scattering cross section,
it is  useful to have analytical expressions for both quantities,
which is however challenging, due to the complicated velocity dependence of the DM annihilation
cross section in the case with resonance.

The Boltzmann equation which governs the evolution of the DM number density
is usually solved analytically by
using the standard freeze-out approximation
\cite{Steigman:2012nb}.
However,
when  DM annihilation takes place near a pole in the cross section,
we cannot use the standard method as
$\langle\sigma v_{\rm rel}\rangle$ does not have a simple analytical form~\cite{Griest:1990kh}.
In Ref.~\cite{Gondolo:1990dk},
it was proposed  to analytically calculate $\langle\sigma v_{\rm rel}\rangle$ using
the $\delta$-function approximation,
if the resonance has a very narrow decay width.
But this method fails in the case where the DM mass is greater
than a half of the mass of the  resonant particle, namely, above the resonance.


In this work,
we present an improved  analytical calculation of
the DM relic density in the case of  resonant DM annihilation  for $s$- and $p$-wave cases and
investigate  the condition for the DM-nuclei scattering cross section to be
above the neutrino background.
We show that
in Higgs-portal type models,
for DM particles with $s$-wave annihilation,
the spin-independent DM-nucleus scattering cross section is proportional to
$\Gamma_{R}/m_{R}$,
the ratio of the decay width and the mass of  $R$.
For a typical DM particle mass $\sim50$~GeV, the condition leads to
$\Gamma_{R}/m_{R} \gtrsim \mathcal{O}(10^{-4})$.
In $p$-wave annihilation case,
the  spin-independent scattering cross section is insensitive to $\Gamma_{R}/m_{R}$,
and
is always above the neutrino background,
as long as the DM particle  is lighter than the top quark.
As an example,
we calculate the spin-independent cross section  both analytically and numerically
in the real singlet DM model with resonant annihilation.
We show that the predicted cross section in this model is always above the
neutrino background.
In order to cover the full parameter space of this model,
the required sensitivity should reach  $\sim 2.6\times10^{-49}~{\rm cm^2}$
for the next generation direct detection experiments.

This paper is organized as follows: In Sec.~2, we outline the freeze-out approximation of Boltzmann equation, and propose an approximate formula for the relic density of a scalar or fermion DM particle which annihilates through an s-channel scalar resonance which has a narrow decay width. In Sec.~3, we analyse the constraint of neutrino backgroud has on DM direct detection experiments on the resonance point. In Sec.~4, we analyse the direct detection of the real singlet DM. Some discussions and conclusions are given in Sec. 5.


\section{DM relic density from resonant annihilation} \label{freeze-out}

The time evolution of the number density $n$ of the DM particle $\chi$ is described by the Boltzmann equation
\begin{eqnarray}\label{11}
\frac{d n}{dt}=-3Hn-\langle\sigma v_{\rm rel}\rangle(n^2-n_{\rm eq}^2),
\end{eqnarray}
where $n_{\rm eq}$ is the equilibrium number density of $\chi$, $H$ is the Hubble parameter, and $\langle\sigma v_{{\rm rel}}\rangle$ is the thermal average of the total annihilation cross section times the relative velocity $v_{{\rm rel}}$ of the annihilating particles. In the non-relativistic case, the thermally averaged cross section can be written as
\begin{eqnarray} \label{12}
\langle \sigma v_{{\rm rel}}\rangle=\frac{x^{3/2}}{2\pi^{1/2}}\int^\infty_0(\sigma v_{{\rm rel}}) v_{{\rm rel}}^2 e^{-xv_{{\rm rel}}^2/4}dv_{{\rm rel}},
\end{eqnarray}
where $x\equiv m_\chi/T$ with $T$ the temperature of the photon in equlibrium and $m_\chi$ the mass of the DM particle. Defining $Y=n/s$ as the comoving density of particle $\chi$ with $s$ the entropy density, Eq.~(\ref{11}) can be rewritten as
\begin{equation} \label{13}
\frac{dY}{dx}=-\sqrt{\frac{\pi g_*}{45}}\frac{M_{\rm pl}m_\chi\langle\sigma v_{{\rm rel}}\rangle}{x^2}(Y^2-Y^2_{{\rm eq}}),
\end{equation}
where $M_{\rm pl}=1.2211\times10^{19}$~GeV is the Plank mass, and
\begin{eqnarray} \label{14}
\sqrt{g_*}=\frac{h_{\rm eff}}{g_{\rm eff}^{1/2}}\left(1+\frac{1}{3}\frac{T}{h_{\rm eff}}\frac{dh_{\rm eff}}{dT}\right),
\end{eqnarray}
where $g_{\rm eff}$ and $h_{\rm eff}$ are the effective relativistic degrees of freedom for entropy and energy density, and
\begin{eqnarray} \label{15}
Y_{\rm eq}=\frac{45}{4\pi^4}\left(\frac{\pi}{8}\right)^{1/2}\frac{g}{h_{\rm eff}}x^{3/2}e^{-x},\ {\rm for}\  x\gg3,
\end{eqnarray}
where $g$ is the internal degrees of freedom of the DM particle $\chi$. The decoupling temperature $x_f$ is defined as the temperature at which the DM particles start to depart from the thermal equilibrium, and the density $Y$ is related to the equilibrium density $Y_{\rm eq}$ by $Y(x_f)\equiv(1+c)Y_{\rm eq}(x_f)$, where $c$ is a constant of order unity. The value of $x_f$ is approximately given by\cite{Scherrer:1985zt}
\begin{eqnarray}
x_f&\approx&\ln[0.038c(c+2)M_{\rm pl}m_\chi g g_{\rm eff}^{-1/2}\langle\sigma v_{{\rm rel}}\rangle]\notag  \\
&\ &-\frac{1}{2}\ln\ln[0.038c(c+2)M_{\rm pl}m_\chi g g_{\rm eff}^{-1/2}\langle\sigma v_{{\rm rel}}\rangle].
\end{eqnarray}
The value of $c$ is usually taken to be one, which leads to a good fit to the numerical solutions of the Boltzmann equation.
The DM number density in the present day $Y_0$ can be obtained by integrating Eq. (\ref{13}) with respect to $x$ in the region $x_f<x<\infty$,
\begin{eqnarray}
\frac{1}{Y_0}&=&\frac{1}{Y(x_f)}+\sqrt{\frac{\pi g_*}{45}}M_{\rm pl}m_\chi\int^{\infty}_{x_f}\frac{\langle\sigma v_{{\rm rel}}\rangle}{x^2}dx\notag  \\
&\approx&\sqrt{\frac{\pi g_*}{45}}M_{\rm pl}m_\chi J_f,
\end{eqnarray}
the function $J_f$ is defined as
\begin{eqnarray} \label{17}
J_f=\int^{\infty}_{x_f}dx\frac{\langle\sigma v_{{\rm rel}}\rangle}{x^2}=\int^{\infty}_{x_f}\frac{dx}{x^2}\frac{x^{3/2}}{2\pi^{1/2}}\int^{\infty}_0dv_{\rm rel}v_{\rm rel}^2(\sigma v_{\rm rel})e^{-xv_{\rm rel}^2/4},
\end{eqnarray}
where we have used the definition of $\langle\sigma v_{{\rm rel}}\rangle$ in Eq.~(\ref{12}). Exchanging the order of the integration in Eq.~(\ref{17}), $J_f$ can be represented as\cite{Griest:1990kh}
\begin{eqnarray} \label{19}
J_f&=&\int^{\infty}_0dv_{\rm rel}\frac{v_{\rm rel}^2(\sigma v_{\rm rel})}{2\pi^{1/2}}\int^{\infty}_{x_f}dxx^{-1/2}e^{-xv_{\rm rel}^2/4} \notag  \\
&=&\int^{\infty}_{0}(\sigma v_{{\rm rel}})v_{\rm rel}{\rm erfc}(\sqrt{x_f}v_{{\rm rel}}/2)dv_{{\rm rel}}.
\end{eqnarray}
The relic density of $\chi$ is obtained from $Y_0$ as
\begin{eqnarray} \label{110}
\Omega h^2=2.755\times10^{8}Y_0\left(\frac{m_\chi}{\rm GeV}\right)\approx2.755\times10^{8}\sqrt{\frac{45}{\pi g_*}}\frac{1}{M_{\rm pl}J_f}\rm GeV^{-1}.
\end{eqnarray}

\subsection{The case of $s$-wave annihilation}
We first consider a real scalar DM particle $\chi$ annihilating into SM particles through exchanging a mediator particle $R$ in $s$-channel. The interaction between $\chi$ and $R$ can be written as $\mathcal{L}\supset\frac{1}{2}\mu\chi\chi R+\mu_HRH^{\dagger}H$, where $\mu$ is a dimensional coupling constant. The term $\mu_HRH^{\dagger}H$ leads to a mixing between $R$ and the SM Higgs boson $H$.
$\chi$ can be stable due to a $Z_2$ symmetry $\chi\leftrightarrow-\chi$. The annihilation proceeds through $s$-wave, the corresponding cross section multiplied by $v_{\rm rel}$ is given by
\begin{equation} \label{21}
\sigma v_{\rm rel}=\frac{2\mu^2}{(s-m_R^2)^2+m_R^2\Gamma_R^2}\frac{\sum_i\Gamma( R^*\rightarrow X_i)}{2m_\chi},
\end{equation}
where $m_R$ and $\Gamma_R$ are the mass and total decay width of the resonance $R$, $s$ is the Mandelstam variable, in the non-relativistic case $s\approx4m_{\chi}^2+m_{\chi}^2v^2_{\rm rel}$, $R^*\rightarrow X_i$ stands for any possible decay mode of $R^*$ and $\sum_i\Gamma(R^*\rightarrow X_i)$ is its total decay width. If $m_\chi$ is close to the resonant point ($\sqrt s\approx m_R\approx2m_\chi$), $\sum_i\Gamma(R^*\rightarrow X_i)$ can be taken as the total decay width $\Gamma_R$ and Eq.~(\ref{21}) can be rewritten as
\begin{equation} \label{22}
\sigma v_{\rm rel}=\frac{\mu^2}{2m_{\chi}^4}\frac{\gamma_R}{(v_{{\rm rel}}^2-\epsilon_R)^2+\gamma_R^2},
\end{equation}
where
\begin{eqnarray}\label{23}
\epsilon_R=\frac{m_R^2-4m_{\chi}^2}{m_{\chi}^2}\ {\rm and}\ \gamma_R=\frac{m_R\Gamma_R}{m_{\chi}^2}.
\end{eqnarray}
From Eq.~(\ref{17}) and Eq.~(\ref{22}), the expression of $J_f$ can be rewritten as
\begin{eqnarray}\label{24}
J_f=\int^{\infty}_{0}\frac{\mu^2}{2m_{\chi}^4}{\rm erfc}(\sqrt{x_f}v_{{\rm rel}}/2)\frac{\gamma_Rv_{{\rm rel}}}{(v_{{\rm rel}}^2-\epsilon_R)^2+\gamma_R^2}dv_{{\rm rel}}.
\end{eqnarray}
There is no analytical expression available for $J_f$. If $\gamma_R\ll1$ and $\gamma_R^2\ll(v_{{\rm rel}}^2-\epsilon_R)^2$, using the relation
\begin{eqnarray}\label{25}
\lim\limits_{\gamma_R\to0}\frac{\gamma_R}{(v_{{\rm rel}}^2-\epsilon_R)^2+\gamma_R^2}=\pi\delta(v_{{\rm rel}}^2-\epsilon_R),
\end{eqnarray}
the value of $J_f$ can be approximated as\cite{Gondolo:1990dk}
\begin{eqnarray}\label{26}
J_f\approx J_f^{d}=\frac{\pi\mu^2}{4m_{\chi}^4}{\rm erfc}(\sqrt{x_f\epsilon_R}/2),\ \  {\rm for} \ \  \epsilon_R > 0\ {\rm and}\ \gamma_R\ll1.
\end{eqnarray}
Note, however that this approximation is only valid for $\epsilon_R>0$.

In this paper we present an improved method to evaluate $J_f$ which is valid for both $\epsilon_R>0$ and $\epsilon_R\leq0$ with a reasonable precision. If $\epsilon_R\geq 0$ and $\gamma_R\ll1$, the integral of Eq.~(\ref{24}) dominates in the narrow region near the point $v_{\rm rel}=\sqrt{\epsilon_R}$. In this region the complementary error function ${\rm erfc}(\sqrt{x_f}v_{{\rm rel}}/2)$ changes very little. We can take ${\rm erfc}(\sqrt{x_f}v_{{\rm rel}}/2)\approx{\rm erfc}(\sqrt{x_f\epsilon_R}/2)$, therefore
\begin{eqnarray}\label{27}
J_f\approx J_f^a&=&\frac{\mu^2\gamma_R}{2m_{\chi}^4}{\rm erfc}(\sqrt{x_f\epsilon_R}/2)\int^{\infty}_{0}\frac{v_{{\rm rel}}}{(v_{{\rm rel}}^2-\epsilon_R)^2+\gamma_R^2}dv_{{\rm rel}}  \notag\\
&=&\frac{\mu^2}{4m_{\chi}^4}{\rm erfc}(\sqrt{x_f\epsilon_R}/2)\left(\frac{\pi}{2}+\arctan\frac{\epsilon_R}{\gamma_R}\right),\ \  {\rm for} \ \  \epsilon_R > 0.
\end{eqnarray}
For $\epsilon_R\gg\gamma_R$, $\arctan(\epsilon_R/\gamma_R)\approx \pi/2$ and $J_f^{a}\approx J_f^{d}$.
Likewise, if $\epsilon_R< 0$ and the absolute value of $\epsilon_R$ approaches zero, the integral of Eq. (\ref{24}) dominates in the region near $v_{\rm rel}=0$ and we can take ${\rm erfc}(\sqrt{x_f}v_{{\rm rel}}/2)\approx1$, then
 \begin{eqnarray}\label{28}
 J_f \approx J_f^a=\frac{\mu^2}{4m_{\chi}^4}\left(\frac{\pi}{2}+\arctan\frac{\epsilon_R}{\gamma_R}\right),\ \  {\rm for}\ \  \epsilon_R < 0.
\end{eqnarray}
In Figure 1(a), we show the differences in the approximate analytical results $J_f^d$, $J_f^a$, and the numerical result $J_f^n$ in a specific case where the parameters are taken as $\mu=1$~GeV, $m_R$=200~GeV, $\sqrt{g_*}=10$, $x_f=20$, and $\Gamma_R=0.001$~GeV. From the figure, if $m_\chi\lesssim100$~GeV, the analytical result $J_f^{a}$ agrees with the numerical result $J_f^{n}$ very well, the relative error is less than 2\%, and the analytical result $J_f^{d}$ can obtain the same precision if $\epsilon_R\gg\gamma_R$ is satisfied. If $m_\chi\gtrsim100$GeV, the approximation of $J_f^{d}$ is no longer valid, but $J_f^{a}$ still agrees with the numerical result well near the resonance point with the relative error is within 12\% (in the region $100~{\rm GeV}\lesssim m_\chi\lesssim100.01$~GeV). The error increases with $m_\chi$ leaving away from the resonance point.

From Eq.~(\ref{110}), (\ref{27}) and (\ref{28}), the relic density of $\chi$ can be represented as
\begin{numcases}{\Omega h^2\approx}\label{32}
2.755\times 10^{8}\sqrt{\frac{45}{\pi g_*}}\frac{4m_{\chi}^4}{\mu^2M_{\rm pl}{\rm erfc}(\sqrt{x_f\epsilon_R/{2}})(\frac{\pi}{2}+\arctan\frac{\epsilon_R}{\gamma_R})}{\rm GeV^{-1}},\ \  {\rm for}\ \   \epsilon_R > 0.   \notag   \\
2.755\times 10^{8}\sqrt{\frac{45}{\pi g_*}}\frac{4m_{\chi}^4}{\mu^2M_{\rm pl}(\frac{\pi}{2}+\arctan\frac{\epsilon_R}{\gamma_R})}{\rm GeV^{-1}},\ \ \ \ \ \ \ \ \ \ \ \ \ \ \ \ \  {\rm for}\ \    \epsilon_R \leq 0.
\end{numcases}

On the resonance point ($m_\chi\approx m_R/2$), we find
\begin{eqnarray}\label{33}
\Omega h^2 \approx2.755\times 10^{8} \sqrt{\frac{ 45}{\pi g_*}}\frac{8m_\chi^4}{\pi\mu^2M_{\rm pl}}{\rm GeV^{-1}}.
\end{eqnarray}
Eq.~(\ref{33}) shows that the relic density is not sensitive to the decay width of $R$ on the resonance point.

If $\chi$ is a complex scalar, the interaction between $\chi$ and $R$ is $\mathcal{L}\supset\mu|\chi|^2R$, the expression for $\sigma v_{\rm rel}$ and relic density is identical to the case of real scalar DM.

\subsection{The case of $p$-wave annihilation}
If $\chi$ is a Dirac particle, the interaction between $\chi$ and $R$ can have the form $\mathcal{L}\supset\lambda_f\overline{\chi}\chi R$ with $\lambda_f$ the coupling constant. The $s$-channel annihilation cross section is a $p$-wave process which is suppressed by $v^2_{\rm rel}$, the cross section is given by
\begin{equation}\label{321}
\sigma v_{\rm rel}=\frac{\lambda_f^2m_{\chi}^2v_{\rm rel}^2}{(s-m_R^2)^2+m_R^2\Gamma_R^2}\frac{\sum_i\Gamma(R^*\rightarrow X_i)}{2m_\chi}.
\end{equation}
Similarly, the expression for $J_f$ is
\begin{eqnarray}\label{322}
J_f=\int^{\infty}_{0}\frac{\lambda_f^2}{4m_{\chi}^2}{\rm erfc}(\sqrt{x_f}v_{{\rm rel}}/2)\frac{\gamma_Rv_{\rm rel}^3}{(v_{{\rm rel}}^2-\epsilon_R)^2+\gamma_R^2}dv_{{\rm rel}}.
\end{eqnarray}
Using the $\delta$-function approximation, one finds
\begin{eqnarray}\label{323}
J_f\approx J_f^{d}=\frac{\pi\epsilon_R\lambda_f^2}{8m_{\chi}^2}{\rm erfc}(\sqrt{x_f\epsilon_R}/2),\ \  {\rm for} \ \  \epsilon_R > 0\ {\rm and}\ \gamma_R\ll1.
\end{eqnarray}
Again this approximation does not apply to the case with $\epsilon_R\leq0$ and $|\epsilon_R|$ approaching zero. The integration of Eq.~(\ref{322}) dominates in the region near the point $v^2_{\rm rel}=\epsilon_R$ if $\gamma_R\ll1$, and the integrand decreases rapidly with $v_{{\rm rel}}^2$ leaving away from $\epsilon_R$. Since the situation we considered is near the resonance point ($\epsilon_R\approx0$), the integral of Eq.~(\ref{24}) can be done in the region $0\lesssim\sqrt{x_f}v_{{\rm rel}}/2\lesssim1$. Using the Taylor expansion of the complementary error function
 \begin{eqnarray}
{\rm erfc}(\sqrt{x_f}v_{{\rm rel}}/2)&=&1-\frac{2}{\sqrt\pi}\left[\sqrt{x_f}v_{\rm rel}/2-\frac{(\sqrt{x_f}v_{\rm rel}/2)^3}{3}+\frac{(\sqrt{x_f}v_{\rm rel}/2)^5}{10}\right.\notag \\
&\ &\left.-\frac{(\sqrt{x_f}v_{\rm rel}/2)^7}{42}+\frac{(\sqrt{x_f}v_{\rm rel}/2)^9}{216}-\cdots\right],
\end{eqnarray}
and retaining the first order term of $v_{\rm rel}$ in the series, $J_f$ can be approximated with a reasonable precision as follows.
\begin{eqnarray}\label{324}
J_f\approx J_f^a&=&\int^{\frac{2}{\sqrt{x_f}}}_{0}\frac{\lambda_f^2\gamma_R}{4m_{\chi}^2}(1-\sqrt{\frac{x_f}{\pi}}v_{\rm rel})\frac{v_{\rm rel}^3}{(v_{{\rm rel}}^2-\epsilon_R)^2+\gamma_R^2}dv_{{\rm rel}} \notag \\
&=&\frac{\lambda_f^2\gamma_R}{8m_{\chi}^2}\Bigg[\ln\sqrt{\frac{(\epsilon_R-4/x_f)^2+\gamma_R^2}{\epsilon_R^2+\gamma_R^2}}
+\frac{\epsilon_R}{\gamma_R}\left(\frac{\pi}{2}+\arctan\frac{\epsilon_R}{\gamma_R}\right)  \notag \\
&-&d\left(\arctan\frac{b+2\sqrt{2/x_f}}{c} -\arctan\frac{b-2\sqrt{2/x_f}}{c}\right)\notag \\
&-&e\ln\frac{a+2b\sqrt{2/x_f}+4/x_f}{a-2b\sqrt{2/x_f}+4/x_f}-\frac{4}{\sqrt\pi}\Bigg],
\end{eqnarray}
where
\begin{eqnarray}
a&=&\sqrt{\epsilon_R^2+\gamma_R^2},\ \ b=\sqrt{a+\epsilon_R},\ \ c=\sqrt{a-\epsilon_R},\notag \\
d&=&\frac{1}{2\gamma_R^2}\sqrt\frac{2x_f}{\pi}b\left(a\epsilon_R+\epsilon_R^2-\gamma_R^2\right),\ \ e=\frac{1}{4\gamma_R^2}\sqrt\frac{2x_f}{\pi}c\left(-a\epsilon_R+\epsilon_R^2-\gamma_R^2\right).
\end{eqnarray}

\begin{figure}[h]\begin{center}
\centering
\subfigure[]{\includegraphics[ width = 2.95in]{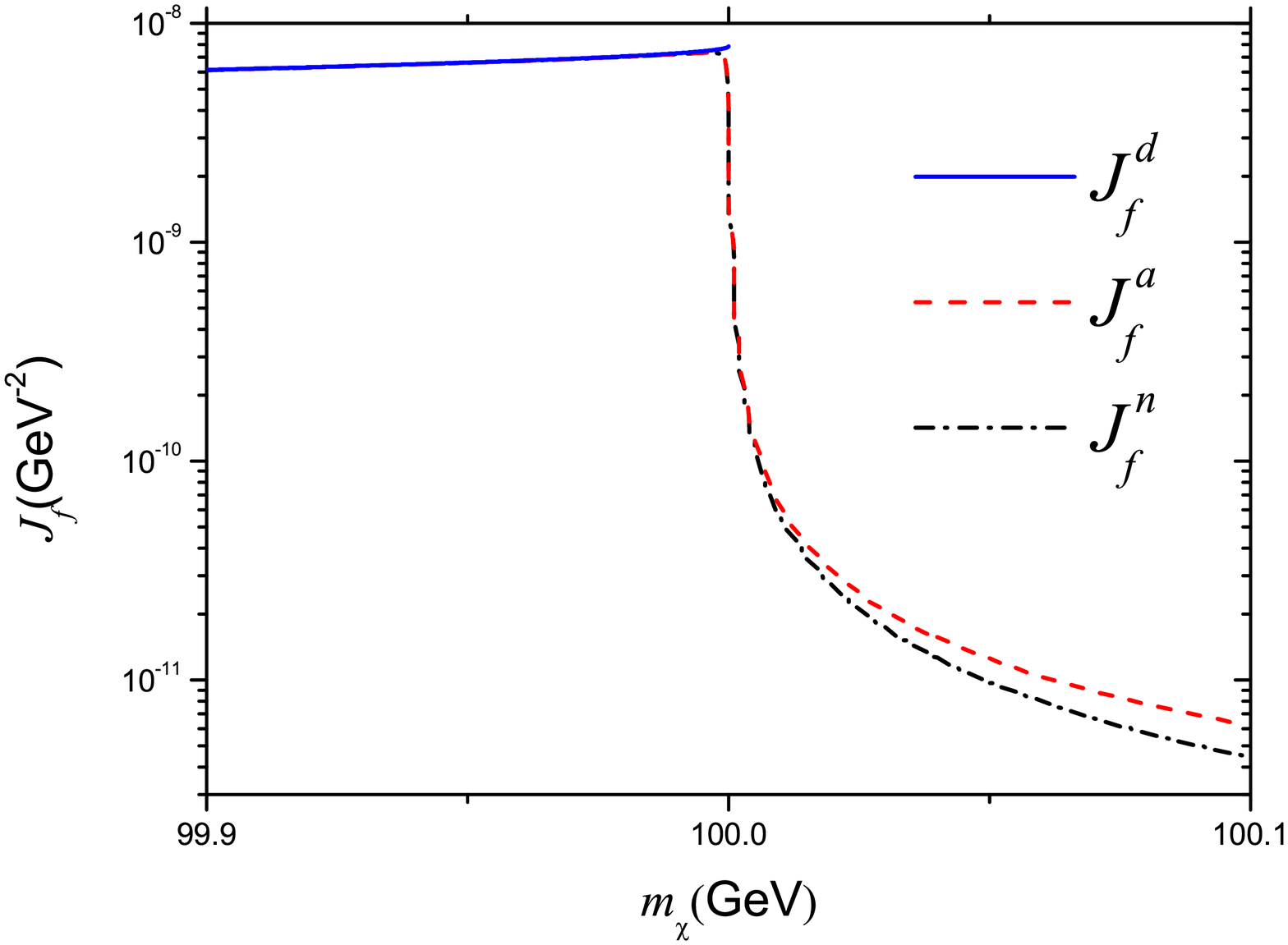}\label{fig1}}
\hspace{0in}
\subfigure[]{
\includegraphics[ width = 2.95in]{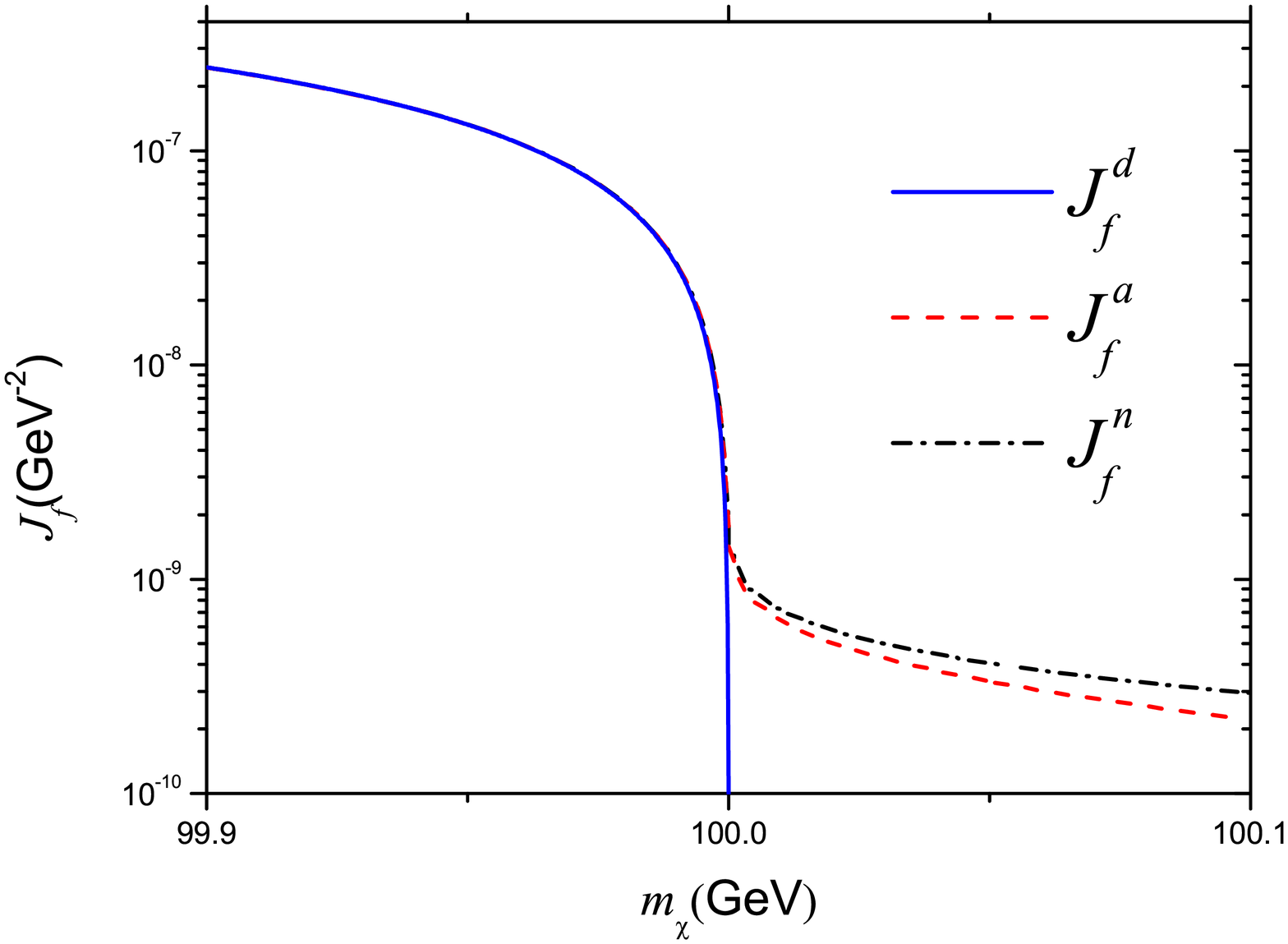}\label{fig2}}
\caption{Analytical approximations $J_f^{d}$, $J_f^{a}$ and numerical result $J_f^{n}$ near the resonance point when $\chi$ is a scalar (a) or fermionic (b) DM particle in a specific case. The parameter values are taken as $\mu=1$~GeV, $\lambda_f=1$, $m_R$=200~GeV, $\sqrt{g_*}$=10, $x_f=20$, and $\Gamma_R=0.001$~GeV.
}
\end{center}
\end{figure}
In Figure 1(b), we show the differences in approximate analytical results $J_f^d$, $J_f^a$, and the numerical result $J_f^n$ in a specific case where the parameters are taken as $\lambda_f=1$, $m_R$=200~GeV, $\sqrt{g_*}=10$, $x_f=20$, and $\Gamma_R=0.001$~GeV. As can be seen from the figure, if $m_\chi\lesssim100$~GeV, the analytical result $J_f^a$ agrees with the numerical result $J_f^n$ well, the relative error is less than 5\%, and the analytical $J_f^d$ can obtain the same precision if $\epsilon_R\gg\gamma_R$ is satisfied. If $m_\chi\gtrsim100$~GeV, $J^d_f$ is no longer valid, but $J_f^a$ still agrees with $J_f^n$ near the resonance point well and the relative error is within 11\% (in the region $100~{\rm~GeV}\lesssim m_\chi\lesssim100.01$~GeV).
On the resonance point, the relic density is given by
\begin{eqnarray}\label{326}
\Omega h^2 \approx 2.755\times10^{8}\sqrt{\frac{ 45}{\pi g_{*}}}\frac{8m_{\chi}^2}{\lambda_f^2M_{\rm Pl}\gamma_R\left[\ln(2x_f^{-1/2}\gamma_R^{-1})-4/\sqrt\pi\right]}{\rm GeV^{-1}}.
\end{eqnarray}
Unlike the $s$-wave case, the relic density is inversely proportional to $\gamma_R$.

If $\chi$ is Majorana fermion, we can write down the Lagrangian of $\chi$ interacting with $R$ as: $\mathcal{L}\supset\frac{1}{2}\lambda_f\overline{\chi}\chi R$, the expression for $\sigma v_{\rm rel}$ and relic density is identical to the case of Dirac DM.

\section{DM direct detection}

Direct detection experiments search for the signal of DM via their interactions with nucleus (for a review, see e.g. ~\cite{Gaitskell:2004gd}). A DM particle can interact with nuclei through $t$-channel scalar $R$ exchange. Since $R$ mixes with the Higgs boson, it can couple to the SM fermions with coupling constant $m_f/v_s$, where $m_f$ is the fermion mass and $v_s$ is a mass scale parameter. The decay width $\Gamma_R$ of the scalar $R$ is given by
\begin{eqnarray} \label{331}
\Gamma_R=\sum_{f}\frac{n_c\eta m_f^2}{8\pi v_s^2}\frac{(m_R^2-4m_f^2)^{\frac{3}{2}}}{m_R^2},
\end{eqnarray}
where $n_c=3$(1) is the number of color for quarks (leptons), $\eta=1$(1/2) for (in)distinguishable final particles.
If $\chi$ is a scalar DM particle, the spin-independent DM-nucleus elastic scattering cross section is given by\cite{Berlin:2014tja}
\begin{equation}\label{34}
\sigma_n^{\rm SI}=\frac{\mu_{\chi n}^2\mu^2}{4\pi m^2_{\chi}m^4_R}f_n^2,
\end{equation}
where $\mu_{\chi n}$ is the DM-nucleon reduced mass $\mu_{\chi n}=m_{\chi}m_{n}/(m_{\chi}+m_{n})$ with $m_n$ is the target nucleus mass. $f_{n}$ stands for the coupling between $R$ and nucleus, which is given by
\begin{equation}\label{35}
\frac{f_n}{m_n}=\sum_{q=u,d,s}f^{(n)}_{Tq}\frac{a_q}{m_q}+\frac{2}{27}f^{(n)}_{TG}\sum_{q=c,b,t}\frac{a_q}{m_q},
\end{equation}
where $ f^{(n)}_{Tu}=0.011,\ f^{(n)}_{Td}=0.0273,$ and $f^{(n)}_{Ts}=0.0447$ \cite{Belanger:2013oya}. The coupling $f^{(n)}_{TG}$ between DM and gluons from heavy quark loops is obtained from $f^{(n)}_{TG}=1-\Sigma_{q=u,d,s}f^{(n)}_{Tq}$, which leads to $f^{(n)}_{TG}= 0.917$. In this case $a_q= m_q/v_s$, then
\begin{eqnarray}\label{38}
\sigma^{\rm SI}_n = \frac{0.02056\mu_{\chi n}^2m_n^2}{\pi v_s^2 m_R^4m_{\chi}^2}\mu^2.
\end{eqnarray}
Making use of Eq.~(\ref{326}), (\ref{331}), and the latest experimental observation $\Omega_{c} h^2 =0.1199\pm0.0027$ \cite{Ade:2013zuv}, we obtain the expression of $\sigma_n^{\rm SI}$ for the DM annihilation into SM fermions through the resonant state $R$
\begin{eqnarray}
\sigma^{\rm SI}_n
\approx\frac{4.12\times10^{-12}\mu_{\chi n}^2m_\chi\gamma_R}{\sqrt{g_{*}}\sum\limits_f n_c\eta m_f^2(m_\chi^2-m_f^2)^{\frac{3}{2}}}.
\end{eqnarray}
The above expression shows that $\sigma^{\rm SI}_n$ is proportional to $\gamma_R$.

For the fermionic DM particle described in section 2.2, it also interacts with nuclei through $t$-channel scalar $R$ exchange.
The spin-independent DM-neucleus elastic scattering cross section is
\begin{eqnarray}\label{39}
\sigma^{\rm SI}_n &=& \frac{0.08224\mu_{\chi n}^2m_n^2}{\pi v_s^2 m_R^4}\lambda_f^2 \notag \\
&\approx&\frac{1.65\times10^{-11}\mu_{\chi n}^2m_\chi}{\sqrt{g_{*}}\left[\ln(2x_f^{-1/2}\gamma_R^{-1})-4/\sqrt\pi\right]\sum\limits_f n_c\eta m_f^2(m_\chi^2-m_f^2)^{\frac{3}{2}}}.
\end{eqnarray}
Compared with the scalar DM case, $\sigma^{\rm SI}_n$ is not sensitive to $\gamma_R$.

For DM direct detection experiments, there is an irreducible background created by the coherent scattering of cosmic neutrinos off target neuclei. This irreducible background is very difficult to be distinguished from the interactions between DM-nuclei scattering, and it can set a limit on the sensitivity of DM direct detection experiments. Due to the neutrino background, the sensitivity of the spin-independent DM-nucleus scattering cross section $\sigma^{\rm SI}_n$ of DM direct detection experiments is limited to $10^{-46}~\rm cm^2\sim10^{-48}~\rm cm^2$, depending on the DM mass\cite{Billard:2013qya,Gutlein:2010tq}. If $\sigma^{\rm SI}_n$ is below the neutrino background, the signal of DM particles can not be reached by DM direct detection experiments.

In Figure 2, we show the relation between $\sigma^{\rm SI}_n$ and $m_\chi$ on the resonance point with different values of $\gamma_R$ when $\chi$ is a scalar (a) or fermion (b). In Figure 2(a), for a typical DM particle mass $\sim50$ GeV, $\sigma^{\rm SI}_n$ is above the neutrino background when the condition $\gamma_R\gtrsim2.2\times10^{-4}$ is satisfied. In Figure 2(b), $\sigma^{\rm SI}_n$ is always above the neutrino background, as long as the DM particle is lighter than the top quark.
\begin{figure}[h]
\begin{center}
\subfigure[]
 {\includegraphics[ width = 3.0in]{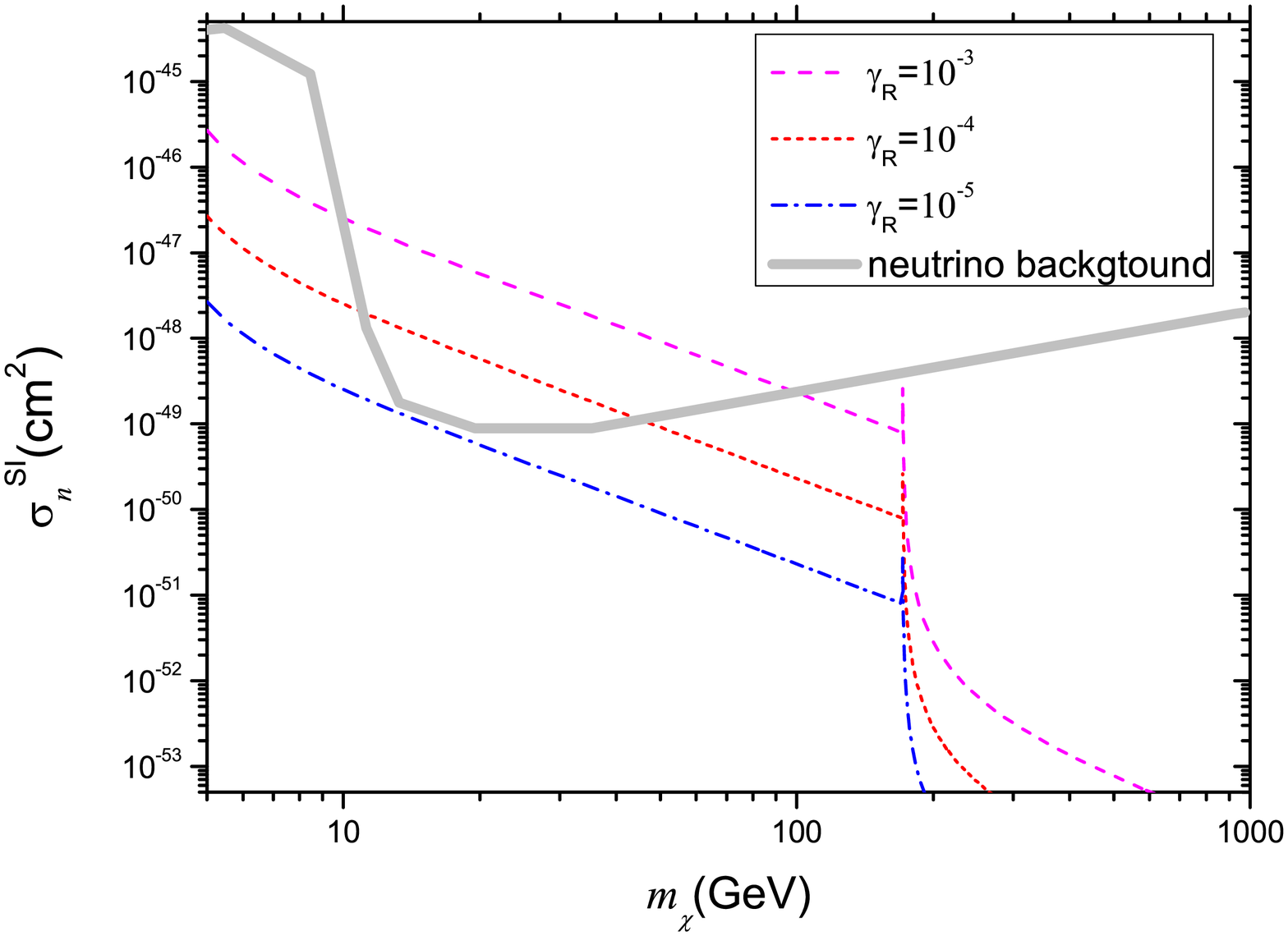}
 \label{fig1}}
\hspace{0\textwidth}
\subfigure[]
 {\includegraphics[ width = 3.0in]{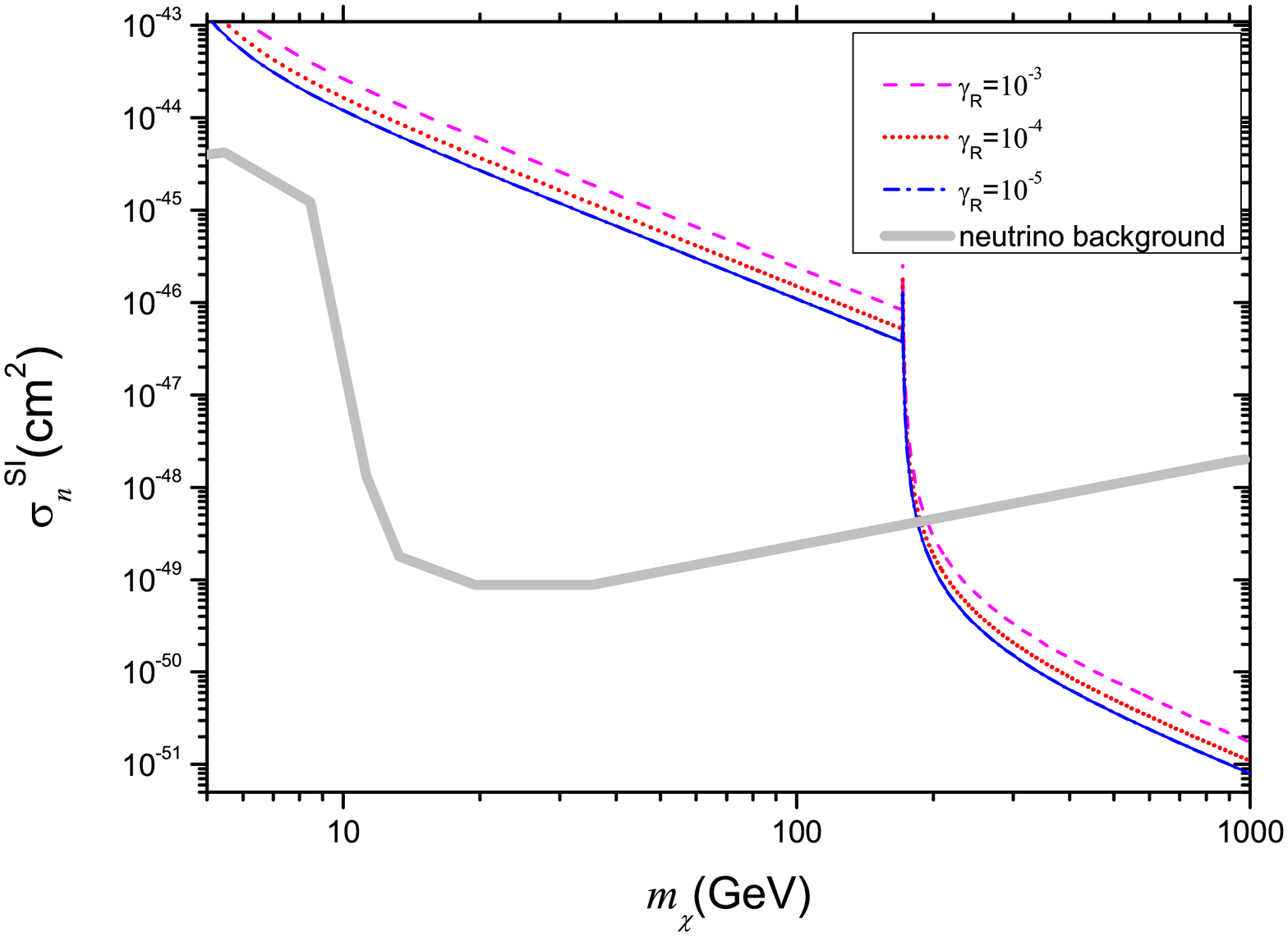}
 \label{fig2}}
\caption{The relation between $\sigma_n^{\rm SI}$ and $m_\chi$ with different $\gamma_R$ for $\chi$ being a scalar (a) or fermion (b) DM particle. We take the parameters value $\sqrt{g_*}=10$, $x_f=20$.
}
\end{center}
\end{figure}

\section{Resonant annihilation in the real singlet dark matter model}

In this section, we consider the resonant annihilation of DM particle in the real singlet DM model\cite{Silveira:1985rk,McDonald:1993ex,Burgess:2000yq,Davoudiasl:2004be,He:2009yd,Gonderinger:2009jp,Mambrini:2011ik,Cline:2012hg,Cline:2013gha,Guo:2010hq}. The Lagrangian of the real singlet DM model is\cite{Silveira:1985rk,Burgess:2000yq}
\begin{eqnarray}\label{41}
 \mathcal{L}=\mathcal{L}_{SM}+\frac{1}{2}\partial_{\mu}D \partial^{\mu}D-\frac{m_0^2}{2}D^2-\frac{\lambda_D}{4}D^4-\lambda_{H} D^2H^{\dagger}H,
\end{eqnarray}
where $\mathcal{L}_{SM}$ is the Lagrangian of SM, $H$ is the SM Higgs doublet. The linear and cubic terms are forbidden due to a discrete $Z_2$ symmetry $D\to-D$. $D$ has a vanishing vacuum expectation value (VEV) to ensure the DM stability. $\lambda_D$ describes the DM self-integration strength which is independent of the DM annihilation. It is clear that the DM-Higgs coupling $\lambda_{H}$ is the only one free parameter to regulate the DM annihilation. After the spontaneous symmetry breaking, one can obtain the DM mass $m_D^2=m_0^2+\lambda_H v_{0}^2$ with the vacuum expectation value $v_{0}=246$ GeV.
In the real singlet dark matter model, the DM annihilation cross section is given by
\begin{eqnarray} \label{42}
\sigma v_{\rm rel} = \frac{8\lambda_{H}^2v_0^2}{(s-m_H^2)^2+\Gamma_H^2m_H^2} \frac{\Gamma_H(\sqrt{s})}{2m_D},
\end{eqnarray}
where $\Gamma_H$ is the total decay width of Higgs which may decay to fermion pairs, gauge boson pairs and the real singlet DM pairs if $m_H>2m_D$\cite{Guo:2010hq}, the value of $\Gamma_H(\sqrt{s})$ is given by
\begin{eqnarray}\label{43}
\Gamma_{H}(\sqrt{s}) & = & \frac{\sum n_c m_f^2 }{8\pi v_{0}^2\eta}
\frac{(s - 4 m_f^2)^{1.5}}{s} + \frac{s^{\frac{3}{2}}}{32 \pi
v_{0}^2} \sqrt{1- \frac{4 m_Z^2}{s}} \left(1-
\frac{4m_{Z}^2}{s}+ \frac{12 m_{Z}^4}{s^2}\right)\notag \\
& + &  \frac{s^{\frac{3}{2}}}{16 \pi v_{0}^2} \sqrt{1- \frac{4
m_W^2}{s}} \left(1- \frac{4m_{W}^2}{s}+ \frac{12
m_{W}^4}{s^2}\right) + \frac{\lambda^2 v_{0}^2}{8 \pi}
\frac{\sqrt{s - 4 m_D^2}}{s} \;
\end{eqnarray}
and $\Gamma_{H}=\Gamma_{H}(\sqrt{s})|_{s=m^2_H}$. Eq.~(\ref{42}) can be written as
\begin{eqnarray}\label{44}
\sigma v_{\rm rel}=\frac{2\lambda_{H}^2v_0^2}{m_D^4}\frac{\gamma_H}{(v_{\rm rel}^2-\epsilon_H)^2+\gamma_H^2},
\end{eqnarray}
where
\begin{eqnarray}\label{45}
\epsilon_H=\frac{m_H^2-4m_D^2}{m_D^2}\ {\rm and}\
\gamma_H=\frac{m_h\Gamma_H(m_H)}{m_D^2}.
\end{eqnarray}

The real singlet DM model is a specific example for the case of $s$-wave annihilation we have discussed, the relic density for the real singlet DM and $\sigma_n^{\rm SI}$ near the resonance point is analogous with Eq.~(\ref{32}) and Eq.~(\ref{38}), and they can be obtained by substituting the parameters $\mu$, $\gamma_R$, and $\epsilon_R$ by 2$\lambda_Hv_0$, $\gamma_H$, and $\epsilon_H$. Figure 3 shows the numerical and analytical value of $\sigma^{\rm SI}_n$, and the upper limits for the spin-independent DM-nucleus cross section from LUX\cite{Akerib:2013tjd} and XENON100\cite{Aprile:2011hi}. In the figure, we find $\sigma^{\rm SI}_n$ is above the neutrino background and it is not excluded by the result of LUX and XENON100 near the resonance point ($m_D=m_H/2$).

\begin{figure}[h]
\centering\includegraphics[width=4.2in]{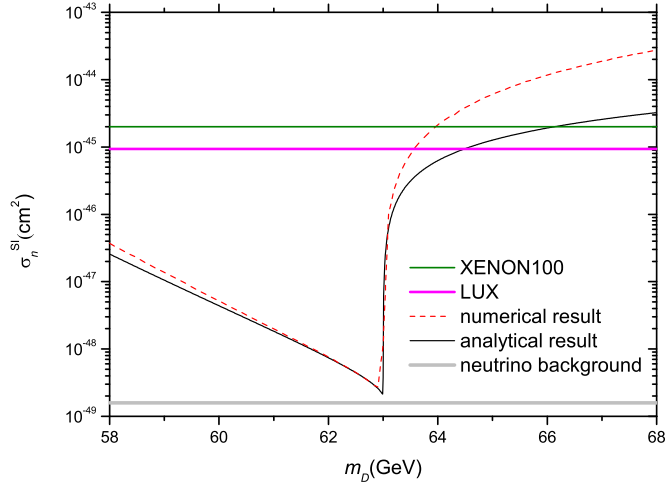}
\caption{The value of $\sigma^{\rm SI}_n$ near the resonance point. Parameter values are $x_f=20$ and $\sqrt{{g_*}}=10$.}
\end{figure}
Currently the strongest upper limits on $\sigma^{\rm SI}_n$ are given by LUX experiment\cite{Akerib:2013tjd} and the next generation of DM direct detection experiments can push the upper bound on $\sigma^{\rm SI}_n$ down to $\sim \ 10^{-47}\ {\rm cm^2}$\cite{Aprile:2012zx}. As the direct detection experiments at present can not measure the $\sigma^{\rm SI}_n$ below $10^{-46}\ {\rm cm^2}$, so we are uncertain of the existing of the real singlet DM near the resonance point. If the future direct detection experiments prove the region near the resonance is excluded, the real singlet DM will be removed from dark matter candidates. In conclusion, if we want to test the singlet dark matter model thoroughly by direct detection, the experiments' ability should reach the minimum value of $\sigma^{\rm SI}_n$, which is about  $2.6\times10^{-49}\ {\rm cm^2}$.

\section{Conclusion}
In summary, we have presented an approximate  analytical expression for
the DM relic density of a scalar or fermionic DM particle which annihilates through an s-channel scalar resonance which has a narrow decay width.
Based on the expression, we have investigated the condition for the DM-nuclei scattering cross section to be above the neutrino background.
It is found that
in Higgs-portal type models,
for DM particles with $s$-wave annihilation,
the spin-independent DM-nucleus scattering cross section is proportional to
$\gamma_R$.
For a typical DM particle mass $\sim50$~GeV, the condition leads to
$\gamma_R \gtrsim \mathcal{O}(10^{-4})$.
In $p$-wave annihilation case,
the  spin-independent scattering cross section is insensitive to $\gamma_R$,
and
is always above the neutrino background,
as long as the DM particle  is lighter than the top quark.
In the real singlet DM model,
$\sigma^{\rm SI}_n$ is always above the
neutrino background.
In order to cover the full parameter space of this model,
the required sensitivity should reach  $\sim 2.6\times10^{-49}~{\rm cm^2}$
for the next generation direct detection experiments.

\section*{Acknowledgement}
This work is supported in part by the National Basic Research Program of China (973 Program) under Grants No.~2010CB833000; the National Nature Science Foundation of China (NSFC) under Grants No.~10975170, No.~10821504, No.~10905084 and No.~11335012; and the Project of Knowledge Innovation Program (PKIP) of the Chinese Academy of Science.
\bibliographystyle{JHEP}
\bibliography{reference}
\end{document}